\documentclass[twocolumn]{revtex4}
\usepackage{amsmath}
\usepackage{graphicx}

\begin{document}

\title{Generalized thermo vacuum state derived by the partial trace method
\thanks{{\small Worked supported by the National Natural Science Foundation
of China under Grant 10775097.}}}
\author{Li-yun Hu$^{1,2}$\thanks{{\small Corresponding author. E-mail:
hlyun2008@126.com }} and Hong-yi Fan$^{1}$}
\affiliation{$^{1}${\small Department of Physics, Shanghai Jiao Tong University, Shanghai
200030, China}\\
$^{2}${\small College of Physics \& Communication Electronics, Jiangxi
Normal University, Nanchang 330022, China}}

\begin{abstract}
{\small By virtue of the technique of integration within an ordered product
(IWOP) of operators we present a new approach for deriving generalized
thermo vacuum state which is simpler in form that the result by using the
Umezawa-Takahashi approach, in this way the thermo field dynamics can be
developed. Applications of the new state are discussed.}


\end{abstract}

\maketitle

\section{Introduction}

In nature every system is immersed in an environment, the problem about the
system interacting with the environment is a hot topic in quantum
information and quantum optics. To describe system-environment evolution
Takahashi-Umezawa introduced thermo field dynamics (TFD) \cite{1,2}, with
which one may convert the calculations of ensemble averages at finite
temperature%
\begin{equation}
\left\langle A\right\rangle =\mathtt{tr}\left( A\rho \right) /Z\left( \beta
\right) ,\text{ }\rho =e^{-\beta H},  \label{1}
\end{equation}%
to equivalent expectation values with a pure state $\left\vert 0(\beta
)\right\rangle $, i.e.,%
\begin{equation}
\left\langle A\right\rangle =\left\langle 0(\beta )\right\vert A\left\vert
0(\beta )\right\rangle ,  \label{2}
\end{equation}%
where $\beta =1/kT$, $k$ is the Boltzmann constant, and $Z\left( \beta
\right) =\mathtt{tr}\rho =\mathtt{tr}e^{-\beta H}$ is the partition
function; $H$ is the system's Hamiltonian. Then how to find the explicit
expression of $\left\vert 0(\beta )\right\rangle ?$ If one expands $%
\left\vert 0(\beta )\right\rangle $ in terms of the energy eigenvector set
of $H$, $\left\vert 0(\beta )\right\rangle =\sum_{n}\left\vert
n\right\rangle f_{n}(\beta ),$ and then substituting it into Eq.(\ref{2}),
which results in $f_{n}^{\ast }(\beta )f_{m}(\beta )=Z^{-1}\left( \beta
\right) e^{-\beta E_{n}}\delta _{n,m}$ (after comparing with Eq.(\ref{1})).
By introducing a fictitious mode, $\left\langle \tilde{n}\right. \left\vert
\tilde{m}\right\rangle =\delta _{n,m},$ then Takahashi-Umezawa obtained
\begin{equation}
\left\vert 0(\beta )\right\rangle =Z^{-1/2}\left( \beta \right)
\sum_{n}e^{-\beta E_{n}/2}\left\vert n,\tilde{n}\right\rangle .  \label{3}
\end{equation}%
Thus the worthwhile convenience in Eq.(\ref{2}) is at the expense of
introducing a fictitious field (or called a tilde-conjugate field, denoted
as operator $\tilde{a}^{\dagger }$) in the extended Hilbert space, i.e., the
original optical field state $\left\vert n\right\rangle $\ in the Hilbert
space $\mathcal{H}$\ is accompanied by a tilde state $\left\vert \tilde{n}%
\right\rangle $\ in $\mathcal{\tilde{H}}$. A similar rule holds for
operators: every Bose annihilation operator $a$\ acting on $\mathcal{H}$\
has an image $\tilde{a}$\ acting on $\mathcal{\tilde{H}}$, $\left[ \tilde{a},%
\tilde{a}^{\dagger }\right] =1$. These operators in $\mathcal{H}$ are
commutative with those in $\mathcal{\tilde{H}}$.

For a harmonic oscillator \ the Hamiltonian is $\hbar \omega a^{\dagger }a,$
$\left\vert n\right\rangle =a^{\dagger n}/\sqrt{n!}\left\vert 0\right\rangle
,$Takahashi-Umezawa obtained the explicit expression of $\left\vert 0(\beta
)\right\rangle $ in doubled Fock space,
\begin{equation}
\left\vert 0(\beta )\right\rangle =\text{sech}\theta \exp \left[ a^{\dagger }%
\tilde{a}^{\dagger }\tanh \theta \right] \left\vert 0\tilde{0}\right\rangle
=S\left( \theta \right) \left\vert 0\tilde{0}\right\rangle ,  \label{4}
\end{equation}%
which is named thermo vacuum state, and $S\left( \theta \right) $ thermo
operator,
\begin{equation}
S\left( \theta \right) \equiv \exp \left[ \theta \left( a^{\dagger }\tilde{a}%
^{\dagger }-a\tilde{a}\right) \right] ,  \label{5}
\end{equation}%
which is similar in form to the a two-mode squeezing operator except for the
tilde mode. $\theta $ is a parameter related to the temperature by $\tanh
\theta =\exp \left( -\frac{\hbar \omega }{2kT}\right) .$

An interesting question thus challenges us: For the Hamiltonian being $%
H=\omega a^{\dagger }a+\kappa ^{\ast }a^{\dagger 2}+\kappa a^{2},$ then what
is the corresponding thermo vacuum state? One may wonder if this question is
worth of paying attention since this $H$ can be diagonalized by the
Bogoliubov transformation as a new harmonic oscillator, correspondingly, the
thermo vacuum state for $H$ can be obtained by acting the same
transformation on $\left\vert 0(\beta )\right\rangle $ in (\ref{4}) (see Eq.
(A9) in the Appendix). To make this issue worthwhile, we emphasize that we
shall adopt a completely new approach to construct thermo vacuum state and
our result is simpler in form than that in Eq. (A10). Our work is arranged
as follows. In Sec. 2 by re-analyzing Eqs. (\ref{1})-(\ref{2}) we shall
introduce a new method (the partial trace method) to find the explicit
expression of $\left\vert 0(\beta )\right\rangle $ in (\ref{4}). Then using
this method, we obtain the expression of $\left\vert 0(\beta )\right\rangle $
in Eq. (\ref{4}) in Sec. 3. For the degenerate parametric amplifier, we
derive a generalized thermal vacuum state $\left\vert \phi (\beta
)\right\rangle $ in Sec. 4. Section 5 is devoting to presenting some
applications of $\left\vert \phi (\beta )\right\rangle $.

\section{The partial trace method}

Following the spirit of TFD, for a density operator $\rho =e^{-\beta
H}/Z\left( \beta \right) $ with Hamiltonian $H$, we can suppose that the
ensemble averages of a system operator $A$ may be calculated as%
\begin{equation}
A=\mathtt{tr}\left( \rho A\right) =\left\langle \psi (\beta )\right\vert
A\left\vert \psi (\beta )\right\rangle ,  \label{6a}
\end{equation}%
where $\left\vert \psi (\beta )\right\rangle $ corresponds to the pure state
in the extended Hilbert space.

Let $\mathtt{Tr}$ denote the trace operation over both the system freedom
(expressed by $\mathtt{tr}$) and the environment freedom by $\widetilde{%
\mathtt{tr}}$, i.e., $\mathtt{Tr}=\mathtt{tr}\widetilde{\mathtt{tr}}$, then
we have%
\begin{eqnarray}
\left\langle \psi (\beta )\right\vert A\left\vert \psi (\beta )\right\rangle
&=&\mathtt{Tr}\left[ A\left\vert \psi (\beta )\right\rangle \left\langle
\psi (\beta )\right\vert \right]   \notag \\
&=&\mathtt{tr}\left[ A\widetilde{\mathtt{tr}}\left\vert \psi (\beta
)\right\rangle \left\langle \psi (\beta )\right\vert \right] .  \label{7}
\end{eqnarray}%
Note that
\begin{equation}
\widetilde{\mathtt{tr}}\left\vert \psi (\beta )\right\rangle \left\langle
\psi (\beta )\right\vert \neq \left\langle \psi (\beta )\right\vert \left.
\psi (\beta )\right\rangle ,  \label{8}
\end{equation}%
since $\left\vert \psi (\beta )\right\rangle $ involves both real mode and
fictitious mode. Comparing Eq.(\ref{7}) with Eq.(\ref{1}) we see
\begin{equation}
\widetilde{\mathtt{tr}}\left\vert \psi (\beta )\right\rangle \left\langle
\psi (\beta )\right\vert =e^{-\beta H}/Z\left( \beta \right) .  \label{9}
\end{equation}%
Eq.(\ref{9}) indicates that, for a given Hamiltonian $H$, if we can find a
density operator\ of pure state $\left\vert \psi (\beta )\right\rangle $\ in
doubled Hilbert space, whose partial trace over the tilde freedom may lead
to density operator $e^{-\beta H}/Z\left( \beta \right) $\ of the system,
then the average value of\textbf{\ }operator $A$ can be calculated as an
equivalent expectation value with a pure state $\left\vert \psi (\beta
)\right\rangle ,$ i.e., $\left\langle A\right\rangle =\mathtt{tr}\left(
Ae^{-\beta H}/Z\left( \beta \right) \right) =\left\langle \psi (\beta
)\right\vert A\left\vert \psi (\beta )\right\rangle $.

In particular, when $H=\hbar \omega a^{\dagger }a,$ a free Bose system, Eq. (%
\ref{9}) becomes
\begin{equation}
\widetilde{\mathtt{tr}}\left\vert 0(\beta )\right\rangle \left\langle
0(\beta )\right\vert =\left( 1-e^{-\beta \hbar \omega }\right) e^{-\beta
\hbar \omega a^{\dagger }a}\equiv \rho _{c},  \label{10}
\end{equation}%
$\rho _{c}$ is the density operator of chaotic field. This equation
enlightens us to have a new approach for deriving $\left\vert 0(\beta
)\right\rangle \colon $ $\left\vert 0(\beta )\right\rangle \left\langle
0(\beta )\right\vert $\ in doubled Hilbert space should be such constructed
that its partial trace over the tilde freedom may lead to density operator $%
\rho _{c}$\textbf{\ }of the system. In the following we shall employ the
technique of integration within an ordered product (IWOP) of operators \cite%
{3,4,5} to realize this goal.

\section{Derivation of $\left\vert 0(\protect\beta )\right\rangle $ in Eq.(%
\protect\ref{4}) via the new approach}

Using the normally ordered expansion formula \cite{6}
\begin{equation}
e^{-\beta \hbar \omega a^{\dagger }a}=\colon \exp \left\{ \left( e^{-\beta
\hbar \omega }-1\right) a^{\dagger }a\right\} \colon ,  \label{11}
\end{equation}%
(where the symbol $\colon \colon $\ denotes the normal ordering form of
operator), and the IWOP technique we have%
\begin{eqnarray}
&&\colon \exp \{\left( e^{-\beta \hbar \omega }-1\right) a^{\dagger
}a\}\colon   \notag \\
&=&\int \frac{d^{2}z}{\pi }\colon e^{-\left\vert z\right\vert ^{2}+z^{\ast
}a^{\dag }e^{-\beta \hbar \omega /2}+zae^{-\beta \hbar \omega /2}-a^{\dag
}a}\colon .  \label{12}
\end{eqnarray}%
Remembering the ordering form of vacuum projector operator $\left\vert
0\right\rangle \left\langle 0\right\vert =\colon e^{-a^{\dag }a}\colon $, we
can rewrite Eq.(\ref{12}) as
\begin{eqnarray}
&&\colon \exp \{\left( e^{-\beta \hbar \omega }-1\right) a^{\dagger
}a\}\colon   \notag \\
&=&\int \frac{d^{2}z}{\pi }e^{z^{\ast }a^{\dag }e^{-\beta \hbar \omega
/2}}\left\vert 0\right\rangle \left\langle 0\right\vert e^{zae^{-\beta \hbar
\omega /2}}\left\langle \tilde{z}\right. \left\vert \tilde{0}\right\rangle
\left\langle \tilde{0}\right. \left\vert \tilde{z}\right\rangle ,  \label{13}
\end{eqnarray}%
where $\left\vert \tilde{z}\right\rangle $ is the coherent state \cite{7,8}
in fictitous mode
\begin{equation}
\text{ \ \ }\left\vert \tilde{z}\right\rangle =\exp \left( z\tilde{a}%
^{\dagger }-z^{\ast }\tilde{a}\right) \left\vert \tilde{0}\right\rangle ,%
\tilde{a}\left\vert \tilde{z}\right\rangle =z\left\vert \tilde{z}%
\right\rangle ,\left\langle \tilde{0}\right. \left\vert \tilde{z}%
\right\rangle =e^{-\left\vert z\right\vert ^{2}/2\text{\ }}.  \label{14}
\end{equation}%
Further, multipling the factor $\left( 1-e^{-\beta \hbar \omega }\right) $
to both sides of Eq.(\ref{13}) and using the completeness of coherent state $%
\int \frac{d^{2}z}{\pi }\left\vert \tilde{z}\right\rangle \left\langle
\tilde{z}\right\vert =1$, we have%
\begin{eqnarray}
&&\left( 1-e^{-\beta \hbar \omega }\right) \times \text{ Eq}.(\ref{13})
\notag \\
&=&\left( 1-e^{-\beta \hbar \omega }\right) \int \frac{d^{2}z}{\pi }\text{ }%
\left\langle \tilde{z}\right\vert e^{z^{\ast }a^{\dag }e^{-\beta \hbar
\omega /2}}\left\vert 0\tilde{0}\right\rangle \left\langle 0\tilde{0}%
\right\vert e^{zae^{-\beta \hbar \omega /2}}\left\vert \tilde{z}%
\right\rangle   \notag \\
&=&\left( 1-e^{-\beta \hbar \omega }\right) \int \frac{d^{2}z}{\pi }\text{ }%
\left\langle \tilde{z}\right\vert e^{a^{\dag }\tilde{a}^{\dag }e^{-\beta
\hbar \omega /2}}\left\vert 0\tilde{0}\right\rangle \left\langle 0\tilde{0}%
\right\vert e^{a\tilde{a}e^{-\beta \hbar \omega /2}}\left\vert \tilde{z}%
\right\rangle   \notag \\
&=&\widetilde{\mathtt{tr}}\left[ 0(\beta )\left\langle 0(\beta )\right\vert %
\right] ,  \label{15}
\end{eqnarray}%
where%
\begin{equation}
\left\vert 0(\beta )\right\rangle =\sqrt{1-e^{-\beta \hbar \omega }}\exp %
\left[ a^{\dagger }\tilde{a}^{\dagger }e^{-\beta \hbar \omega /2}\right]
\left\vert 0\tilde{0}\right\rangle ,  \label{16}
\end{equation}%
which is the same as Eq.(\ref{4}). Thus, according to Eq.(\ref{10}), from
the chaotic field operator we have derived the thermo vacuum state, this is
a new approach, which has been overlooked in the literature before.

\section{Generalized thermo vacuum state $\left\vert \protect\phi \left(
\protect\beta \right) \right\rangle $}

Now, we consider a degenerate parametric amplifier whose Hamiltonian is
\begin{equation}
H=\omega a^{\dagger }a+\kappa ^{\ast }a^{\dagger 2}+\kappa a^{2},  \label{17}
\end{equation}%
whose normalized density operator $\rho $ is%
\begin{equation}
\rho \left( \mathtt{tr}e^{-\beta H}\right) =e^{-\beta H}=e^{-\beta \left(
\omega a^{\dagger }a+\kappa ^{\ast }a^{\dagger 2}+\kappa a^{2}\right) }.
\label{18}
\end{equation}

Recalling that $\frac{1}{2}\left( a^{\dagger }a+\frac{1}{2}\right) ,\frac{1}{%
2}a^{\dagger 2}$ and $\frac{1}{2}a^{2}$ obey the SU(1,1) Lie algebra, thus
we can derive a generalized identity of operator \cite{9,10} as follows:%
\begin{eqnarray}
&&\exp \left[ fa^{\dagger }a+ga^{\dagger 2}+ka^{2}\right]   \notag \\
&=&e^{-f/2}e^{\frac{ga^{\dagger 2}}{\mathcal{D}\coth \mathcal{D}-f}%
}e^{\left( a^{\dagger }a+\frac{1}{2}\right) \ln \frac{\mathcal{D}\text{sech}%
\mathcal{D}}{\mathcal{D}-f\tanh \mathcal{D}}}e^{\frac{ka^{2}}{\mathcal{D}%
\coth \mathcal{D}-f}},  \label{19}
\end{eqnarray}%
where we have set $\mathcal{D}^{2}=f^{2}-4kg.$ Thus Comparing Eq.(\ref{18})
with Eq.(\ref{19}) we can recast Eq.(\ref{18}) into the following form%
\begin{equation}
\left( \mathtt{tr}e^{-\beta H}\right) \rho =\sqrt{\lambda e^{\beta \omega }}%
\exp \left[ E^{\ast }a^{\dagger 2}\right] \exp \left[ a^{\dagger }a\ln
\lambda \right] \exp \left[ Ea^{2}\right] ,  \label{20}
\end{equation}%
where we have set%
\begin{eqnarray}
D^{2} &=&\omega ^{2}-4\left\vert \kappa \right\vert ^{2},  \notag \\
\lambda  &=&\frac{D}{\omega \sinh \beta D+D\cosh \beta D},  \label{21} \\
E &=&\frac{-\lambda }{D}\kappa \sinh \beta D.  \notag
\end{eqnarray}
Further, using the formula in Eqs.(\ref{11}) and (\ref{13}), we have%
\begin{eqnarray}
&&\left( \mathtt{tr}e^{-\beta H}\right) \rho   \notag \\
&=&\sqrt{\lambda e^{\beta \omega }}\exp \left[ E^{\ast }a^{\dagger 2}\right]
\colon \exp \left\{ \left( \lambda -1\right) a^{\dagger }a\right\} \colon
\exp \left[ Ea^{2}\right]   \notag \\
&=&\sqrt{\lambda e^{\beta \omega }}\int \frac{d^{2}z}{\pi }e^{E^{\ast
}a^{\dagger 2}+\sqrt{\lambda }z^{\ast }a^{\dag }}\left\vert 0\right\rangle
\left\langle 0\right\vert e^{Ea^{2}+\sqrt{\lambda }za}\left\langle \tilde{z}%
\right. \left\vert \tilde{0}\right\rangle \left\langle \tilde{0}\right.
\left\vert \tilde{z}\right\rangle   \notag \\
&=&\sqrt{\lambda e^{\beta \omega }}\int \frac{d^{2}z}{\pi }\left\langle
\tilde{z}\right\vert e^{E^{\ast }a^{\dagger 2}+\sqrt{\lambda }z^{\ast
}a^{\dag }}\left\vert 0\tilde{0}\right\rangle \left\langle 0\tilde{0}\right.
e^{Ea^{2}+\sqrt{\lambda }za}\left\vert \tilde{z}\right\rangle   \notag \\
&=&\sqrt{\lambda e^{\beta \omega }}\widetilde{\mathtt{tr}}\left[ e^{E^{\ast
}a^{\dagger 2}+\sqrt{\lambda }a^{\dag }\tilde{a}^{\dag }}\left\vert 0\tilde{0%
}\right\rangle \left\langle 0\tilde{0}\right. e^{Ea^{2}+\sqrt{\lambda }a%
\tilde{a}}\right]   \notag \\
&\equiv &\left( \mathtt{tr}e^{-\beta H}\right) \widetilde{\mathtt{tr}}\left[
\left\vert \phi \left( \beta \right) \right\rangle \left\langle \phi \left(
\beta \right) \right\vert \right] ,  \label{22}
\end{eqnarray}%
which indicates that the pure state in doubled Fock space for Hamiltonian in
(\ref{17}) can be considered as%
\begin{equation}
\left\vert \phi \left( \beta \right) \right\rangle =\sqrt{\frac{\lambda
^{1/2}e^{\beta \omega /2}}{Z\left( \beta \right) }}e^{E^{\ast }a^{\dagger 2}+%
\sqrt{\lambda }a^{\dag }\tilde{a}^{\dag }}\left\vert 0\tilde{0}\right\rangle
,  \label{23}
\end{equation}%
where the partition function $Z\left( \beta \right) $ is determined by%
\begin{equation}
Z\left( \beta \right) =\mathtt{tr}e^{-\beta H}=\mathtt{tr}\left\{ \sqrt{%
\lambda e^{\beta \omega }}e^{E^{\ast }a^{\dagger 2}}\colon e^{\left( \lambda
-1\right) a^{\dagger }a}\colon e^{Ea^{2}}\right\} .  \label{24}
\end{equation}%
Using $\int \frac{d^{2}z}{\pi }\left\vert z\right\rangle \left\langle
z\right\vert =1$ and the integral formula \cite{11}
\begin{eqnarray}
&&\int \frac{d^{2}z}{\pi }\exp \left( \zeta \left\vert z\right\vert ^{2}+\xi
z+\eta z^{\ast }+fz^{2}+gz^{\ast 2}\right)   \notag \\
&=&\frac{1}{\sqrt{\zeta ^{2}-4fg}}\exp \left[ \frac{-\zeta \xi \eta +\xi
^{2}g+\eta ^{2}f}{\zeta ^{2}-4fg}\right] ,  \label{25}
\end{eqnarray}%
whose convergent condition is Re$\left( \zeta \pm f\pm g\right) <0,\ $Re$%
\left( \frac{\zeta ^{2}-4fg}{\zeta \pm f\pm g}\right) <0,$ we can get
\begin{equation}
Z\left( \beta \right) =\sqrt{\frac{\lambda e^{\beta \omega }}{\left(
1-\lambda \right) ^{2}-4\left\vert E\right\vert ^{2}}}=\frac{e^{\beta \omega
/2}}{2\sinh \left( \beta D/2\right) }.  \label{27}
\end{equation}%
Thus the normalized state for Eq.(\ref{18}) in doubled Fock space is given
by
\begin{equation}
\left\vert \phi \left( \beta \right) \right\rangle =\sqrt{2\lambda
^{1/2}\sinh \left( \beta D/2\right) }e^{E^{\ast }a^{\dagger 2}+\sqrt{\lambda
}a^{\dag }\tilde{a}^{\dag }}\left\vert 0\tilde{0}\right\rangle ,  \label{28}
\end{equation}%
and the internal energy of system is
\begin{equation}
\left\langle H\right\rangle _{e}=-\frac{\partial }{\partial \beta }\ln
Z\left( \beta \right) =\frac{D\coth \left( \beta D/2\right) -\omega }{2},
\label{29}
\end{equation}%
which leads to the distribution of entropy
\begin{eqnarray}
S &=&-k\mathtt{tr}\left( \rho \ln \rho \right) =\frac{1}{T}\left\langle
H\right\rangle _{e}+k\ln Z\left( \beta \right)   \notag \\
&=&\frac{D}{2T}\coth \left( \beta D/2\right) -k\ln \left[ 2\sinh \left(
\beta D/2\right) \right] .  \label{30}
\end{eqnarray}%
In particular, when $\kappa =0,$ leading to $D=\omega ,$ so Eq.(\ref{28})
just reduces to $\left\vert 0(\beta )\right\rangle $ with $\omega
\rightarrow \hbar \omega $, and Eqs.(\ref{29}) and (\ref{30}) become $\frac{%
\omega }{2}\left( \coth \left( \beta \omega /2\right) -1\right) $ and $\frac{%
\omega }{2T}\coth \left( \beta \omega /2\right) -k\ln \left[ 2\sinh \left(
\beta \omega /2\right) \right] ,$ respectively, as expected \cite{12}. Thus
by virtue of the technique of IWOP we can display the partial trace method
to deduce the pure state representation for some new density operators of
light field at finite temperature.

\section{Applications of generalized thermo vacuum state}

\subsection{Internal energy distribution of the system}

As an application of Eq.(\ref{28}), we can evaluate the each term's
contribution to energy in Hamiltonian. Based on the idea from Eq.(\ref{1})
to (\ref{2}), the system operator $A$ can be calculated as $\left\langle
A\right\rangle _{e}=\left\langle \phi \left( \beta \right) \right\vert
A\left\vert \phi \left( \beta \right) \right\rangle .$ Thus uisng the
completeness of coherent state and the integral formula Eq.(\ref{25}) as
well as noticing $\left\langle \phi \left( \beta \right) \right. \left\vert
\phi \left( \beta \right) \right\rangle =1$, $\left( 1-\lambda \right)
^{2}-4\left\vert E\right\vert ^{2}=4\lambda \sinh ^{2}\left( \beta
D/2\right) ,$ then we have%
\begin{eqnarray}
\left\langle \omega a^{\dagger }a\right\rangle _{e} &=&\omega \left\langle
\phi \left( \beta \right) \right\vert \left( aa^{\dagger }-1\right)
\left\vert \phi \left( \beta \right) \right\rangle   \notag \\
&=&2\omega \lambda ^{1/2}\sinh \left( \beta D/2\right) \frac{\partial }{%
\partial \lambda }  \notag \\
&&\times \int \frac{d^{2}z}{\pi }e^{-\left( 1-\lambda \right) \left\vert
z\right\vert ^{2}+Ez^{2}+E^{\ast }z^{\ast 2}}-\omega   \notag \\
&=&\lambda ^{1/2}\frac{\partial }{\partial \lambda }\frac{2\omega \sinh
\left( \beta D/2\right) }{\sqrt{\left( 1-\lambda \right) ^{2}-4\left\vert
E\right\vert ^{2}}}-\omega   \notag \\
&=&\frac{\omega }{2}\left( \frac{\omega }{D}\coth \beta D/2-1\right) ,
\label{31}
\end{eqnarray}%
and%
\begin{eqnarray}
\left\langle \kappa ^{\ast }a^{\dagger 2}\right\rangle _{e} &=&2\kappa
^{\ast }\lambda ^{1/2}\sinh \left( \beta D/2\right) \frac{\partial }{%
\partial E^{\ast }}\frac{1}{\sqrt{\left( 1-\lambda \right) ^{2}-4EE^{\ast }}}
\notag \\
&=&-\frac{\left\vert \kappa \right\vert ^{2}}{D}\allowbreak \coth \beta D/2,
\label{32}
\end{eqnarray}%
as well as%
\begin{equation}
\left\langle \kappa a^{2}\right\rangle _{e}=-\frac{\left\vert \kappa
\right\vert ^{2}}{D}\allowbreak \coth \beta D/2.  \label{33}
\end{equation}%
From Eqs.(\ref{32}) and (\ref{33}) we see that the two items ($\kappa ^{\ast
}a^{\dagger 2}$ and $\kappa a^{2})$ have the same energy contributions to
the system, as expected. Combing Eqs.(\ref{31})-(\ref{33}) we can also check
Eq.(\ref{29}).

\subsection{Wigner function and quantum tomogram}

The Wigner function plays an important role in studying quantum optics and
quantum statistics \cite{13,14}. It gives the most analogous description of
quantum mechanics in the phase space to classical statistical mechanics of
Hamilton systems and is also a useful measure for studying the nonclassical
features of quantum states. In addition, the Wigner function can be
reconstructed by measuring several quadratures $P\left( \hat{x}_{\theta }=%
\hat{x}\cos \theta +\hat{p}\sin \theta \right) $ with a homodyne detection
and then applying an inverse Radon transform---quantum homodyne tomography
\cite{15}. Using Eq.(\ref{27}) one can calculate conveniently the Wigner
function and quantum tomogram. Recalling that the single-mode Wigner
operator $\Delta \left( z\right) $ in coherent state representation is given
by \cite{16,17}
\begin{equation}
\Delta \left( \alpha \right) =e^{2\left\vert \alpha \right\vert ^{2}}\int
\frac{d^{2}z}{\pi ^{2}}\left\vert z\right\rangle \left\langle -z\right\vert
e^{-2\left( z\alpha ^{\ast }-z^{\ast }\alpha \right) },  \label{34}
\end{equation}%
thus the Wigner function is
\begin{eqnarray}
W\left( \alpha \right)  &=&\left\langle \phi \left( \beta \right)
\right\vert \Delta \left( \alpha \right) \left\vert \phi \left( \beta
\right) \right\rangle   \notag \\
&=&e^{2\left\vert \alpha \right\vert ^{2}}\int \frac{d^{2}z}{\pi ^{2}}%
\left\langle \phi \left( \beta \right) \right. \left\vert z\right\rangle
\left\langle -z\right. \left\vert \phi \left( \beta \right) \right\rangle
e^{-2\left( z\alpha ^{\ast }-z^{\ast }\alpha \right) }  \notag \\
&=&\frac{\tanh \left( \beta D/2\right) }{\pi }e^{-\frac{2}{D}\left[ \omega
\left\vert \alpha \right\vert ^{2}+\left( \kappa \alpha ^{2}+\kappa ^{\ast
}\alpha ^{\ast 2}\right) \right] \tanh \left( \beta D/2\right) },  \label{35}
\end{eqnarray}%
where we have noticed $\left( 1+\lambda \right) ^{2}-4\left\vert
E\right\vert ^{2}=\frac{4D\cosh ^{2}\beta D/2}{\omega \sinh \beta D+D\cosh
\beta D}$ and used the integral formula (\ref{25}). In particular, when $%
\kappa =0,$ Eq.(\ref{35}) reduces to
\begin{equation}
W\left( \alpha \right) =\frac{\tanh \left( \beta D/2\right) }{\pi }\exp
\left\{ -2\left\vert \alpha \right\vert ^{2}\tanh \left( \beta D/2\right)
\right\} ,  \label{36}
\end{equation}%
which is just the Wigner function of thermo vacuum state $\left\vert 0\left(
\beta \right) \right\rangle $.

On the other hand, we can derive the tomography (Radon transform of Wigner
function) of the system by using Eq.(\ref{27}). Recalling that, for
single-mode case, the Radon transform of the Wigner operator is just a pure
state density operator \cite{18},%
\begin{equation}
\int \delta \left( q-fq^{\prime }-gp^{\prime }\right) \Delta \left( \alpha
\right) \mathtt{d}q^{\prime }\mathtt{d}p^{\prime }=\left\vert q\right\rangle
_{f,g\text{ }f,g}\left\langle q\right\vert ,  \label{37}
\end{equation}%
where $\alpha =(q+\mathtt{i}p)/\sqrt{2},$ and $\left( f,g\right) $ are real,%
\begin{equation}
\left\vert q\right\rangle _{f,g}=C\exp \left[ \frac{\sqrt{2}}{A}qa^{\dag }-%
\frac{e^{\mathtt{i}2\varphi }}{2}a^{\dag 2}\right] \left\vert 0\right\rangle
,  \label{38}
\end{equation}%
and $C=\left[ \pi \left( f^{2}+g^{2}\right) \right] ^{-1/4}\exp
\{-q^{2}/[2(f^{2}+g^{2})]\},A=f-\mathtt{i}g=\sqrt{f^{2}+g^{2}}e^{-\mathtt{i}%
\varphi }.$ Eq.(\ref{38}) is named as the intermediate coordinate-momentum
representation \cite{18}. From Eq.(\ref{37}) and Eq.(\ref{27}) it then
follows that the tomogram can be calculated as%
\begin{equation}
\mathcal{R}\left( q\right) _{f,g}\equiv \left\langle \phi \left( \beta
\right) \right. \left\vert q\right\rangle _{f,g\text{ }f,g}\left\langle
q\right. \left\vert \phi \left( \beta \right) \right\rangle =\int \frac{%
d^{2}z}{\pi }\left\vert _{f,g}\left\langle q\right\vert \left\langle \tilde{z%
}\right. \left\vert \phi \left( \beta \right) \right\rangle \right\vert ^{2}.
\label{39}
\end{equation}%
Then submitting Eqs.(\ref{38}) and (\ref{27}) into Eq.(\ref{39}), we obtain%
\begin{eqnarray}
\mathcal{R}\left( q\right) _{f,g} &=&\frac{2\sinh \left( \beta D/2\right) }{%
C^{-2}\sqrt{\lambda +\left\vert G\right\vert ^{2}/\lambda }}  \notag \\
&&\times \exp \left\{ 2q^{2}\left[ \frac{1-\lambda \text{Re}G^{-1}}{%
\left\vert A\right\vert ^{2}\left( \lambda +\left\vert G\right\vert
^{2}/\lambda \right) }+\text{Re}\frac{2E}{A^{2}G}\right] \right\} ,
\label{40}
\end{eqnarray}%
where we have used Eq.(\ref{25}) and set $G=1+2e^{\mathtt{i}2\varphi }E$.
Eq.(\ref{40}) is the positive-definite tomogram, as expected. As far as we
are concerned, this result has not been reported in the literature before.

In sum, by virtue of the technique of integration within an ordered product
(IWOP) of operators we have presented a new approach for deriving
generalized thermo vacuum state which is simpler in form that the result by
using the Umezawa-Takahashi approach, in this way the thermo field dynamics
can be developed.

\textbf{Appendix:}

As a comparison of our new approach with the usual way of deriving thermo
vacuum state in TFD theory, in this appendix, we shall derive the explicit
expression of $\left\vert \phi \left( \beta \right) \right\rangle $ by
diagonalizing Hamiltonian (\ref{17}). For this purpose, we introduce two
unitary operators: one is a single mode squeezing operator,%
\begin{equation}
S=\exp \left( \frac{\nu }{\mu }\frac{a^{\dag 2}}{2}\right) \exp \left[
\left( a^{\dag }a+\frac{1}{2}\right) \ln \frac{1}{\mu }\right] \exp \left( -%
\frac{\nu }{\mu }\frac{a^{2}}{2}\right) ,  \tag{A1}
\end{equation}%
where $\mu $ and $\nu $\ are squeezing parameters satisfying the
unitary-modulate condition $\mu ^{2}-\nu ^{2}=1$; and the other is a
rotational operator, $R=\exp \left( \frac{i\phi }{2}a^{\dagger }a\right) ,$
which lead to the following transformations,%
\begin{align}
SaS^{\dag }& =\mu a-\nu a^{\dag },\text{ }Sa^{\dag }S^{\dag }=\mu a^{\dag
}-\nu a,  \tag{A2} \\
S^{\dag }aS& =\mu a+\nu a^{\dag },\text{ }S^{\dag }a^{\dag }S=\mu a^{\dag
}+\nu a,  \tag{A3}
\end{align}%
and%
\begin{equation}
RaR^{\dag }=ae^{-\frac{i\phi }{2}},\text{ }Ra^{\dag }R^{\dag }=a^{\dag }e^{%
\frac{i\phi }{2}}.  \tag{A4}
\end{equation}%
Thus, under the unitary transform $SR$, we have (setting $\kappa =\left\vert
\kappa \right\vert e^{i\phi }$)%
\begin{align}
H^{\prime }& =SRHR^{\dag }S^{\dag }=\omega Sa^{\dagger }aS^{\dag
}+\left\vert \kappa \right\vert Sa^{\dagger 2}S^{\dag }+\left\vert \kappa
\right\vert Sa^{2}S^{\dag }  \notag \\
& =\left( \omega \mu ^{2}+\omega \nu ^{2}-4\left\vert \kappa \right\vert \mu
\nu \right) a^{\dag }a+\left( \omega \nu -2\left\vert \kappa \right\vert \mu
\right) \nu   \notag \\
& +\left( \left\vert \kappa \right\vert \mu ^{2}+\left\vert \kappa
\right\vert \nu ^{2}-\omega \mu \nu \right) \left( a^{\dag 2}+a^{2}\right) .
\tag{A5}
\end{align}%
In order to diagonalize Eq.(A5), noticing $\mu ^{2}-\nu ^{2}=1$ and making $%
\left\vert \kappa \right\vert \left( \mu ^{2}+\nu ^{2}\right) -\omega \mu
\nu =0,$ whose solution is given by
\begin{equation}
\mu ^{2}=\frac{\omega }{2\omega ^{\prime }}+\frac{1}{2},\nu ^{2}=\allowbreak
\frac{\omega }{2\omega ^{\prime }}-\frac{1}{2},\omega ^{\prime }=\sqrt{%
\omega ^{2}-4\left\vert \kappa \right\vert ^{2}}.  \tag{A6}
\end{equation}%
then Eq.(A5) becomes
\begin{equation}
H^{\prime }=\omega ^{\prime }\left( a^{\dagger }a+\frac{1}{2}\right) -\frac{1%
}{2}\omega .  \tag{A7}
\end{equation}%
i.e., the diagonalization of Hamiltonian is completed.

According to Eq.(\ref{16}), the thermal vacuum\ state corresponding to
density operator $\rho ^{\prime }=e^{-\beta H^{\prime }}/\mathtt{tr}\left(
e^{-\beta H^{\prime }}\right) =e^{-\beta \omega ^{\prime }a^{\dagger }a}/%
\mathtt{tr}\left( e^{-\beta \omega ^{\prime }a^{\dagger }a}\right) $ is
given by%
\begin{equation}
\left\vert 0(\beta )\right\rangle =\sqrt{1-e^{-\beta \omega ^{\prime }}}\exp %
\left[ a^{\dagger }\tilde{a}^{\dagger }e^{-\beta \omega ^{\prime }/2}\right]
\left\vert 0\tilde{0}\right\rangle .  \tag{A8}
\end{equation}%
Thus the generalized thermal vacuum state is
\begin{align}
\left\vert \phi ^{\prime }(\beta )\right\rangle & =R^{\dag }S^{\dag
}\left\vert 0(\beta )\right\rangle   \notag \\
& =\sqrt{1-e^{-\beta \omega ^{\prime }}}R^{\dag }S^{\dag }\exp \left[
a^{\dagger }\tilde{a}^{\dagger }e^{-\beta \omega ^{\prime }/2}\right]
\left\vert 0\tilde{0}\right\rangle .  \tag{A9}
\end{align}%
Using the transformation in (A3), (A4) and noticing Eq.(A1) as well as $%
\frac{\nu }{\mu }=\sqrt{\frac{\omega -\omega ^{\prime }}{\omega +\omega
^{\prime }}}$, we can finally put Eq.(A9) into the following form%
\begin{align}
\left\vert \phi ^{\prime }(\beta )\right\rangle & =\sqrt{\left( 1-e^{-\beta
\omega ^{\prime }}\right) /\mu }\exp \left[ \frac{1}{\mu }e^{-\left( \beta
\omega ^{\prime }+i\phi \right) /2}a^{\dag }\tilde{a}^{\dagger }\right.
\notag \\
& \left. -\frac{\nu e^{-i\phi }}{2\mu }a^{\dag 2}+\frac{\nu e^{-\beta \omega
^{\prime }}}{2\mu }\tilde{a}^{\dagger 2}\right] \left\vert 0\tilde{0}%
\right\rangle .  \tag{A10}
\end{align}%
Comparing Eq.(A10) with Eq.(\ref{28}), we see that Eq.(\ref{28}) is simpler
in form than that in Eq.(A10).

\textbf{ACKNOWLEDGEMENT:} Worked supported by the National Natural Science
Foundation of China under Grant 10775097 and 10874174. E-mail:
hlyun2008@126.com.

\end{document}